%% file: root.tex
\documentclass[submission,copyright,creativecommons]{eptcs}
\providecommand{\event}{F-IDE 2018}
\usepackage{breakurl}

\usepackage[utf8]{inputenc}
\usepackage{hyperref}
\usepackage{ifthen}
\usepackage[english]{babel}
\usepackage{graphicx}
\usepackage{color}
\usepackage{isabelle,isabellesym}

\isabellestyle{literal}

\isadroptag{theory}

\definecolor{linkcolor}{rgb}{0,0,0.5}
\hypersetup{colorlinks=true,linkcolor=linkcolor,citecolor=linkcolor,filecolor=linkcolor,urlcolor=linkcolor,pdfpagelabels}

\renewcommand{\isastyletxt}{\isastyletext}
\renewcommand{\isadigit}[1]{\isamath{#1}}

\newcommand{\secref}[1]{\S\ref{#1}}
\newcommand{\figref}[1]{fig.~\ref{#1}}
\newcommand{\Figref}[1]{Fig.~\ref{#1}}

\begin{document}

\title{Isabelle/jEdit as IDE for Domain-specific Formal Languages and
  Informal Text Documents}
\author{Makarius Wenzel \\ \url{https://sketis.net}}
\def\titlerunning{Isabelle/jEdit as Formal IDE}
\def\authorrunning{M. Wenzel}
\maketitle

\begin{abstract}

Isabelle/jEdit is the main application of the Prover IDE (PIDE) framework and
the default user-interface of Isabelle, but it is not limited to theorem
proving. This paper explores possibilities to use it as a general IDE for
formal languages that are defined in user-space, and embedded into informal
text documents. It covers overall document structure with auxiliary files and
document antiquotations, formal text delimiters and markers for interpretation
(via control symbols). The ultimate question behind this: How far can we
stretch a plain text editor like jEdit in order to support semantic text
processing, with support by the underlying PIDE framework?

\end{abstract}

\input{Paper.tex}

\bibliographystyle{eptcs}
\bibliography{root}

\end{document}

%% file: Paper.tex
\begin{isabellebody}%
\setisabellecontext{Paper}%
\isadelimtheory
\isanewline
\isanewline
\endisadelimtheory
\isatagtheory
\isacommand{theory}\isamarkupfalse%
\ Paper\isanewline
\ \ \isakeyword{imports}\ Main\isanewline
\isakeyword{begin}%
\endisatagtheory
{\isafoldtheory}%
\isadelimtheory
\endisadelimtheory
\isadelimdocument
\endisadelimdocument
\isatagdocument
\isamarkupsection{Introduction%
}
\isamarkuptrue%
\endisatagdocument
{\isafolddocument}%
\isadelimdocument
\endisadelimdocument
\begin{isamarkuptext}%
Isabelle is a well-known system for interactive theorem proving\footnote{See
  \url{https://isabelle.in.tum.de/website-Isabelle2018} for the official release
  Isabelle2018 (August 2018).} and jEdit a text editor written in
  Java\footnote{\url{http://jedit.org}}. Isabelle/jEdit is a separate application on its
  own \cite{isabelle-jedit-manual}, based on Isabelle and jEdit: it is a
  particular front-end for the Prover IDE (PIDE) framework, which has emerged
  in the past decade \cite{Wenzel:2010,Wenzel:2012:UITP-EPTCS,Wenzel:2014:UITP-EPTCS,Wenzel:2014:ITP-PIDE}. Another PIDE front-end
  is
  Isabelle/VSCode\footnote{\url{https://marketplace.visualstudio.com/items?itemName=makarius.Isabelle2018}},
  but that is still somewhat experimental and needs to catch up several years
  of development.

  Officially, the main purpose of Isabelle/jEdit is to support the user in
  reading and writing \emph{proof documents}, with continuous checking by the
  prover in the background. This is illustrated by a small example in
  \figref{fig:isabelle-jedit}, which is the first entry in the
  \emph{Documentation} panel: it uses Isabelle/HOL as logical basis and as a
  library of tools for specifications and proofs. In bigger applications, e.g.
  from \emph{The Archive of Formal Proofs} (AFP)\footnote{\url{https://www.isa-afp.org}},
  the overall document may consist of hundreds of source files, with a typical
  size of of 50--500\,KB each.

  \begin{figure}[h!]
  \centering
  \includegraphics[width=0.95\textwidth]{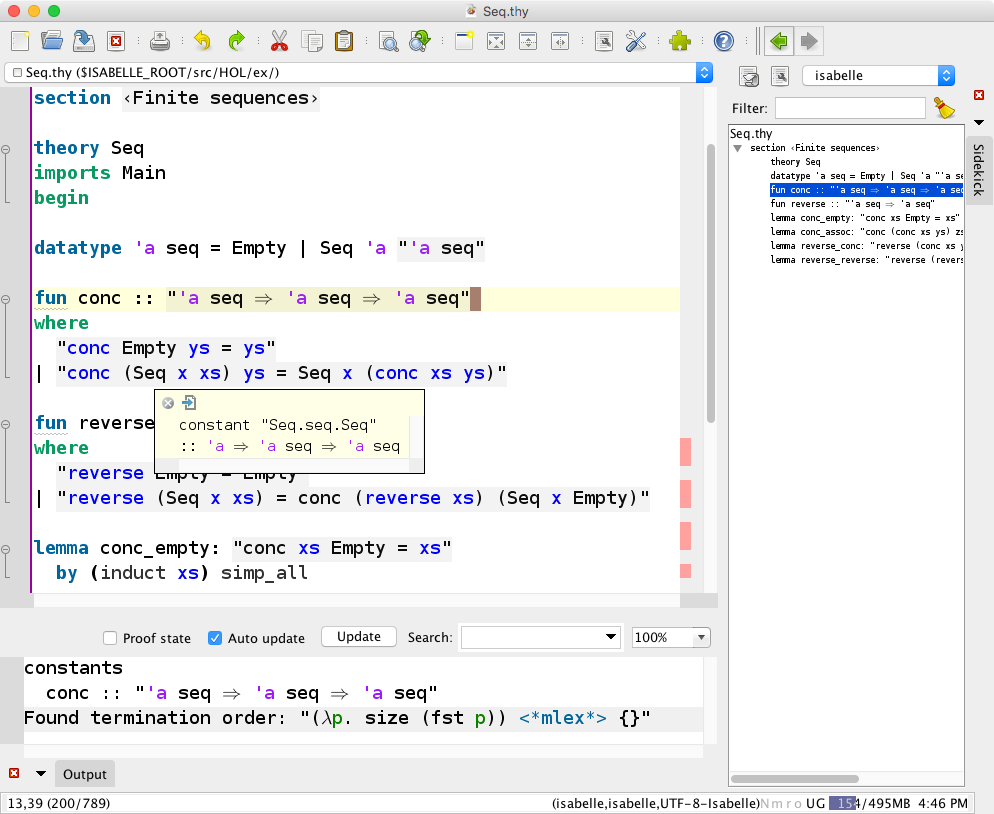}
  \caption{The Isabelle/jEdit Prover IDE with example proof document}
  \label{fig:isabelle-jedit}
  \end{figure}

  Another example document is \emph{this paper} itself.\footnote{See
  \url{https://bitbucket.org/makarius/fide2018/src} for the Mercurial repository
  and note that Isabelle/jEdit can open raw file URLs directly, e.g.
  \url{https://bitbucket.org/makarius/fide2018/raw/tip/Paper.thy} or
  \url{https://bitbucket.org/makarius/fide2018/raw/tip/document/root.bib} with
  semantic IDE support even for BibTeX.} Isabelle documents can be rendered
  with PDF-{\LaTeX} already since 1999, and the world has seen reports,
  articles, books, and theses produced by Isabelle without taking much notice.
  Thus Isabelle could be understood as a formal version of {\LaTeX}, but with
  more explicit structure, better syntax, and substantial IDE support.

  Isabelle documents may contain both formal and informal text, with section
  headings, bullet lists, and paragraphs of plain words. New sub-languages can
  be defined by incorporating suitable Isabelle/ML modules into the theory
  context. The default setup of Isabelle/Pure and Isabelle/HOL already
  provides a wealth of languages for types, terms, propositions, definitions,
  theorem statements, proof text (in Isar), proof methods (in Eisbach) etc.

  A notable example beyond formal logic is the \emph{rail} language for railroad
  syntax diagrams: e.g. see the grammar specification for the \isa{\isacommand{find{\isacharunderscore}theorems}}
  command in the manual \cite[\S3.4.1]{isabelle-jedit-manual} and its
  source in the IDE (\figref{fig:rail}). The implementation consists of a
  small module in Isabelle/ML to parse rail expressions and report the
  recognized structure as PIDE
  markup\footnote{\isatt{{\char`\$}ISABELLE{\char`\_}HOME/\discretionary{}{}{}src/\discretionary{}{}{}Pure/\discretionary{}{}{}Tools/\discretionary{}{}{}rail.ML}}, together with a style
  file for {\LaTeX}.\footnote{\isatt{{\char`\$}ISABELLE{\char`\_}HOME/\discretionary{}{}{}lib/\discretionary{}{}{}texinputs/\discretionary{}{}{}railsetup.sty}}

  \begin{figure}[h!]
  \centering
  \includegraphics[width=0.95\textwidth]{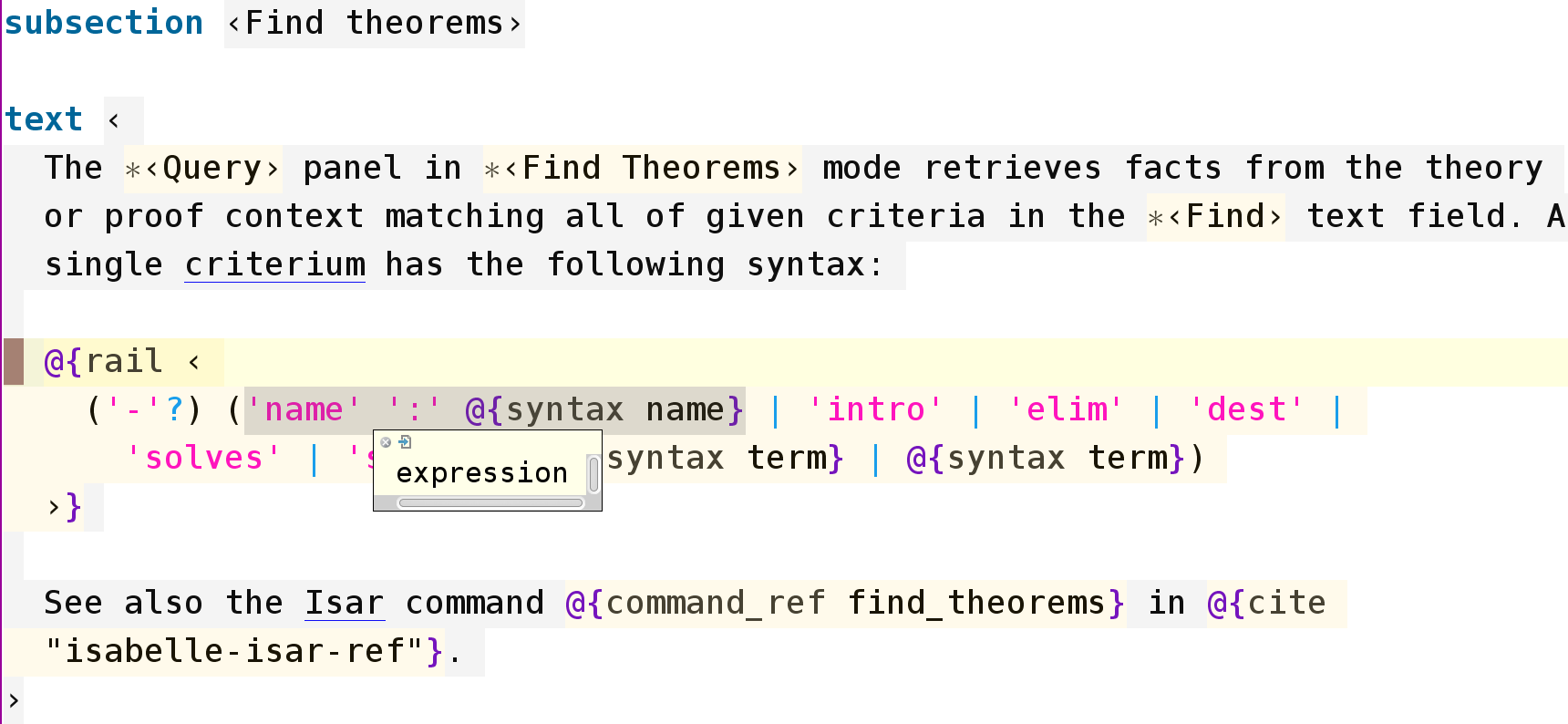}
  \caption{PIDE document view with antiquotation for railroad syntax diagram}
  \label{fig:rail}
  \end{figure}

  \medskip Subsequently, we shall explore further uses of Isabelle/jEdit as a
  general-purpose IDE. \secref{sec:document-structure} gives a systematic
  overview of PIDE document structure, as it is technically managed by
  Isabelle. This may provide some clues to fit other tools into the PIDE
  framework. \secref{sec:spark} illustrates interactive PIDE document
  processing by the example of a SPARK/Ada verification environment from the
  Isabelle library. \secref{sec:document-text} explains document text
  structure, as a combination of formal and informal content.
  \secref{sec:bibtex} presents an example of other file formats managed by the
  IDE: BibTeX databases and citations within PIDE documents.%
\end{isamarkuptext}\isamarkuptrue%
\isadelimdocument
\endisadelimdocument
\isatagdocument
\isamarkupsection{PIDE Document Structure \label{sec:document-structure}%
}
\isamarkuptrue%
\endisatagdocument
{\isafolddocument}%
\isadelimdocument
\endisadelimdocument
\begin{isamarkuptext}%
The structuring principles for PIDE documents have emerged over decades,
  driven by the demands of interactive proof development in Isabelle. The
  resulting status-quo may now be re-used for general documents with mixed
  languages, with some degree of scalability.%
\end{isamarkuptext}\isamarkuptrue%
\isadelimdocument
\endisadelimdocument
\isatagdocument
\isamarkupparagraph{Project directories%
}
\isamarkuptrue%
\endisatagdocument
{\isafolddocument}%
\isadelimdocument
\endisadelimdocument
\begin{isamarkuptext}%
refer to locations in the file-system with session \isatt{ROOT} files (for
  session specifications) and \isatt{ROOTS} files (for lists of project
  sub-directories). Isabelle/jEdit and the Isabelle build tool provide option
  \isatt{{\char`\-}d} to specify additional project directories, the default is the Isabelle
  distribution itself. A common add-on is The Archive of Formal Proofs (AFP).%
\end{isamarkuptext}\isamarkuptrue%
\isadelimdocument
\endisadelimdocument
\isatagdocument
\isamarkupparagraph{Sessions%
}
\isamarkuptrue%
\endisatagdocument
{\isafolddocument}%
\isadelimdocument
\endisadelimdocument
\begin{isamarkuptext}%
form an acyclic graph within a project directory. Sessions can be based on
  the special session \isatt{Pure}, which is the start of the bootstrap process, or
  on other sessions as a single \emph{parent session} or multiple \emph{import
  sessions}. The session graph restricted to the parent relation forms a tree:
  it corresponds to the bottom-up construction of \emph{session images}, i.e.
  persistent store of the Isabelle/ML state as ``dumped world''; these can be
  extended, but not merged. In contrast, session imports merge the name space
  of source files from other sessions, but the resulting session image needs
  to process such side-imports once again, to store them persistently.

  Session names need to be globally unique wrt. the active project
  directories. Notable example sessions are: \isatt{Pure} (Isabelle bootstrap
  environment), \isatt{HOL} (main Isabelle/HOL), \isatt{HOL{\char`\-}Analysis} (a rather big
  Isabelle/HOL library with classic analysis),
  \isatt{Ordinary{\char`\_}Differential{\char`\_}Equations} (an AFP entry that formalizes
  differential equations on top of classic analysis).

  Session content is specified in the session \isatt{ROOT} entry via \textbf{options},
  \textbf{theories} (end points for the reachable import graph), and
  \textbf{document\_files} (headers, styles, graphics etc. for PDF-{\LaTeX} output).%
\end{isamarkuptext}\isamarkuptrue%
\isadelimdocument
\endisadelimdocument
\isatagdocument
\isamarkupparagraph{Theories%
}
\isamarkuptrue%
\endisatagdocument
{\isafolddocument}%
\isadelimdocument
\endisadelimdocument
\begin{isamarkuptext}%
form an acyclic graph within a session. The theory name is qualified
  according to the session name, e.g. \isatt{HOL.List}, but could be global in
  exceptional situations, e.g. \isatt{Pure} (instead of \isatt{Pure.Pure}), \isatt{Main}
  (instead of \isatt{HOL.Main}).

  A theory header is of the form \isa{\isacommand{theory}\ A\ \isakeyword{imports}\ B\isactrlsub {\isadigit{1}}\ {\isasymdots}\ B\isactrlsub n\ \isakeyword{begin}} and
  refers to the theory base name and its \emph{parent theories}. Imports either
  refer to other sessions (session-qualified theory name) or to \isatt{.thy} files
  within the session base directory (relative file name without the
  extension).

  Semantically, a theory is a container for arbitrary \emph{theory data} defined
  in Isabelle/ML, together with \isa{extend} and \isa{merge} operations to propagate
  it along the theory import graph. For example, theory \isa{A} above begins with
  an extension of all data from theories \isa{B\isactrlsub {\isadigit{1}}{\isacharcomma}\ {\isasymdots}{\isacharcomma}\ B\isactrlsub n} merged in a canonical
  order; its body may augment the data further. This is motivated by the
  contents of logical signatures, e.g. types and constants declared in various
  theories become available in the extension due to the \emph{monotonicity} of the
  logical environment. Another example is the Isabelle/ML toplevel environment
  within the theory: ML types, value, signatures, structures, functors are
  propagated monotonically by merging the corresponding name spaces.

  In other words, Isabelle theories organize ``worksheets'' with user-defined
  data, which is propagated monotonically along the foundational order of the
  acyclic relation of imports; there is no support for mutual theory
  dependencies.%
\end{isamarkuptext}\isamarkuptrue%
\isadelimdocument
\endisadelimdocument
\isatagdocument
\isamarkupparagraph{Commands%
}
\isamarkuptrue%
\endisatagdocument
{\isafolddocument}%
\isadelimdocument
\endisadelimdocument
\begin{isamarkuptext}%
form the main body of a theory as sequence of semantic updates on theory
  data. Commands are defined within the theory body by updating a special data
  slot, but command names need to be declared beforehand as \isa{\isakeyword{keywords}} in the
  theory header; this enables the Prover IDE to parse theory structure without
  semantic evaluation.

  For example, the command \isa{\isacommand{definition}} is defined in theory \isatt{Pure} and
  defines a logical constant, based on an existing term from the current
  theory context:%
\end{isamarkuptext}\isamarkuptrue%
\ \ \ \ \ \ \ \ \isacommand{definition}\isamarkupfalse%
\ foo\ {\isacharcolon}{\isacharcolon}\ {\isacartoucheopen}nat\ {\isasymRightarrow}\ nat{\isacartoucheclose}\ \isakeyword{where}\ {\isacartoucheopen}foo\ n\ {\isasymequiv}\ {\isadigit{2}}\ {\isacharasterisk}\ n\ {\isacharplus}\ {\isadigit{3}}{\isacartoucheclose}%
\begin{isamarkuptext}%
As another example, the command \isa{\isacommand{ML}} is defined in the bootstrap process
  before theory \isatt{Pure} and allows to augment the ML environment (as data
  within the current theory) by ML toplevel declarations (types, values,
  signatures, structures etc.). This is the \emph{meta language}, it allows to
  access the implementation of the logic from the outside (with references to
  the formal context):%
\end{isamarkuptext}\isamarkuptrue%
\isadelimML
\ \ \ \ \ \ \ \ %
\endisadelimML
\isatagML
\isacommand{ML}\isamarkupfalse%
\ {\isacartoucheopen}\isanewline
\ \ \ \ \ \ \ \ \ \ val\ foo{\isacharunderscore}def\ {\isacharequal}\ {\isacharat}{\isacharbraceleft}thm\ foo{\isacharunderscore}def{\isacharbraceright}{\isacharsemicolon}\isanewline
\ \ \ \ \ \ \ \ \ \ val\ lhs\ {\isacharequal}\ Thm{\isachardot}lhs{\isacharunderscore}of\ foo{\isacharunderscore}def{\isacharsemicolon}\isanewline
\ \ \ \ \ \ \ \ \ \ val\ rhs\ {\isacharequal}\ Thm{\isachardot}rhs{\isacharunderscore}of\ foo{\isacharunderscore}def{\isacharsemicolon}\isanewline
\ \ \ \ \ \ \ \ \ \ \isactrlassert \ {\isacharparenleft}Thm{\isachardot}typ{\isacharunderscore}of{\isacharunderscore}cterm\ lhs\ {\isacharequal}\ \isactrltyp {\isasymopen}nat{\isasymclose}{\isacharparenright}{\isacharsemicolon}\isanewline
\ \ \ \ \ \ \ \ {\isacartoucheclose}%
\endisatagML
{\isafoldML}%
\isadelimML
\endisadelimML
\begin{isamarkuptext}%
Any command is free to define its own concrete syntax, within the token
  language of \emph{outer syntax} of Isabelle theories, but excluding the keywords
  of other commands. This includes various quotations, e.g. single-quoted or
  double-quoted string literals and \emph{cartouches}, which are text literals
  with balanced quotes that are outside of most other languages in existence:
  \isa{{\isasymopen}text{\isasymclose}}.

  Quotation allows a command to refer to a nested sub-language, and apply its
  own parsing and semantic processing on it, independently of the theory
  syntax. The semantic language context may be stored as theory data. Examples
  for that are again \isa{\isacommand{definition}} (for the embedded term language of
  Isabelle), and \isa{\isacommand{ML}} (for the ML environment that maps names to entities of
  the ML world).%
\end{isamarkuptext}\isamarkuptrue%
\isadelimdocument
\endisadelimdocument
\isatagdocument
\isamarkupparagraph{Auxiliary files%
}
\isamarkuptrue%
\endisatagdocument
{\isafolddocument}%
\isadelimdocument
\endisadelimdocument
\begin{isamarkuptext}%
occur as arguments of special \emph{load commands} within a theory.
Syntactically, the argument is a file path specification, relatively to the
location of the enclosing theory file. Semantically, the argument is the
text content of that file. The Prover IDE manages the physical file within
the editor, and treats its current content as add-on to the load command
within the theory body. The prover only sees the current content, ignorant
of dependency management and edits.

For example, the command \isa{\isacommand{ML{\isacharunderscore}file}} (defined in theory \isatt{Pure}) is
semantically equivalent to \isa{\isacommand{ML}}, but \isa{\isacommand{ML{\isacharunderscore}file}} uses the ML text from the
given \isatt{.ML} file. This improves scalability and usability, e.g. the Prover
IDE can manage the auxiliary file in its own editing mode, with different
syntax rules than for the enclosing \isatt{.thy} file.

\bigskip In summary, tool builders have two main possibilities to include other
languages within a theory body:

\begin{enumerate}%
\item as a regular command with literal text argument, e.g. \isa{\isacommand{ML}\ {\isacartoucheopen}val\ a\ {\isacharequal}\ {\isadigit{1}}{\isacartoucheclose}}

\item as a load command with file path argument, e.g. \isa{\isacommand{ML{\isacharunderscore}file}\ {\isacartoucheopen}a{\isachardot}ML{\isacartoucheclose}}%
\end{enumerate}

\noindent All other document structure merely helps the user and the IDE to organize
large projects, to manage file-system structure and name spaces of theories.

\medskip The commands \isa{\isacommand{ML}} and \isa{\isacommand{ML{\isacharunderscore}file}} have served here as canonical example
for embedded languages within the Isabelle theory environment, but this
language happens to be the main implementation and extension language of
Isabelle itself. It is used to declare new theory data slots and to define
commands operating over the content. This resembles the approach of
Smalltalk: a highly dynamic and extensible environment (with IDE) that is
used to bootstrap itself.

In fact, the Isabelle Prover IDE may be applied to Isabelle/Pure itself by
opening the file \isatt{{\char`\$}ISABELLE{\char`\_}HOME/\discretionary{}{}{}src/\discretionary{}{}{}Pure/\discretionary{}{}{}ROOT.ML} and exploring its uses
of \isa{\isacommand{ML{\isacharunderscore}file}}. Note that \isatt{ROOT.ML} files are treated as special bootstrap
theories in the Prover IDE: this allows to bootstrap pure ML projects within
Isabelle/jEdit. In addition, there are some special tricks to load a fresh
copy of \isatt{Pure} into an already running \isatt{Pure} session.

The special treatment of \isatt{ROOT.ML} files can used to bootstrap other
projects in Standard ML as well, independently of the Isabelle/ML
environment. I have occasionally demonstrated that for the HOL4 theorem
prover\footnote{\url{https://sketis.net/2015/hol4-workshop-at-cade-25}} and
Metitarski\footnote{\url{https://sketis.net/2016/isabellepide-as-ide-for-ml-2}}.%
\end{isamarkuptext}\isamarkuptrue%
\isadelimdocument
\endisadelimdocument
\isatagdocument
\isamarkupsection{Example: Isabelle/HOL-SPARK \label{sec:spark}%
}
\isamarkuptrue%
\endisatagdocument
{\isafolddocument}%
\isadelimdocument
\endisadelimdocument
\begin{isamarkuptext}%
Isabelle/HOL-SPARK is a SPARK/Ada verification environment by Stefan
  Berghofer (secunet Security Networks AG). Technically it consists of a few
  sessions within the project directory of the official Isabelle distribution:
  \isatt{HOL{\char`\-}SPARK}, \isatt{HOL{\char`\-}SPARK{\char`\-}Examples}, \isatt{HOL{\char`\-}SPARK{\char`\-}Manual} (with full PDF
  \cite{isabelle-hol-spark-manual}). As Isabelle/jEdit implicitly operates
  on the union of all sessions from the active project directories, we merely
  need to open any of the example theories and can start working after a few
  seconds of processing library imports: progress can be observed in the
  \emph{Theories} panel. There is no need to build session images beforehand in
  the old-fashioned way, as described in the manual.

  \begin{figure}[h!]
  \centering
  \includegraphics[width=0.95\textwidth]{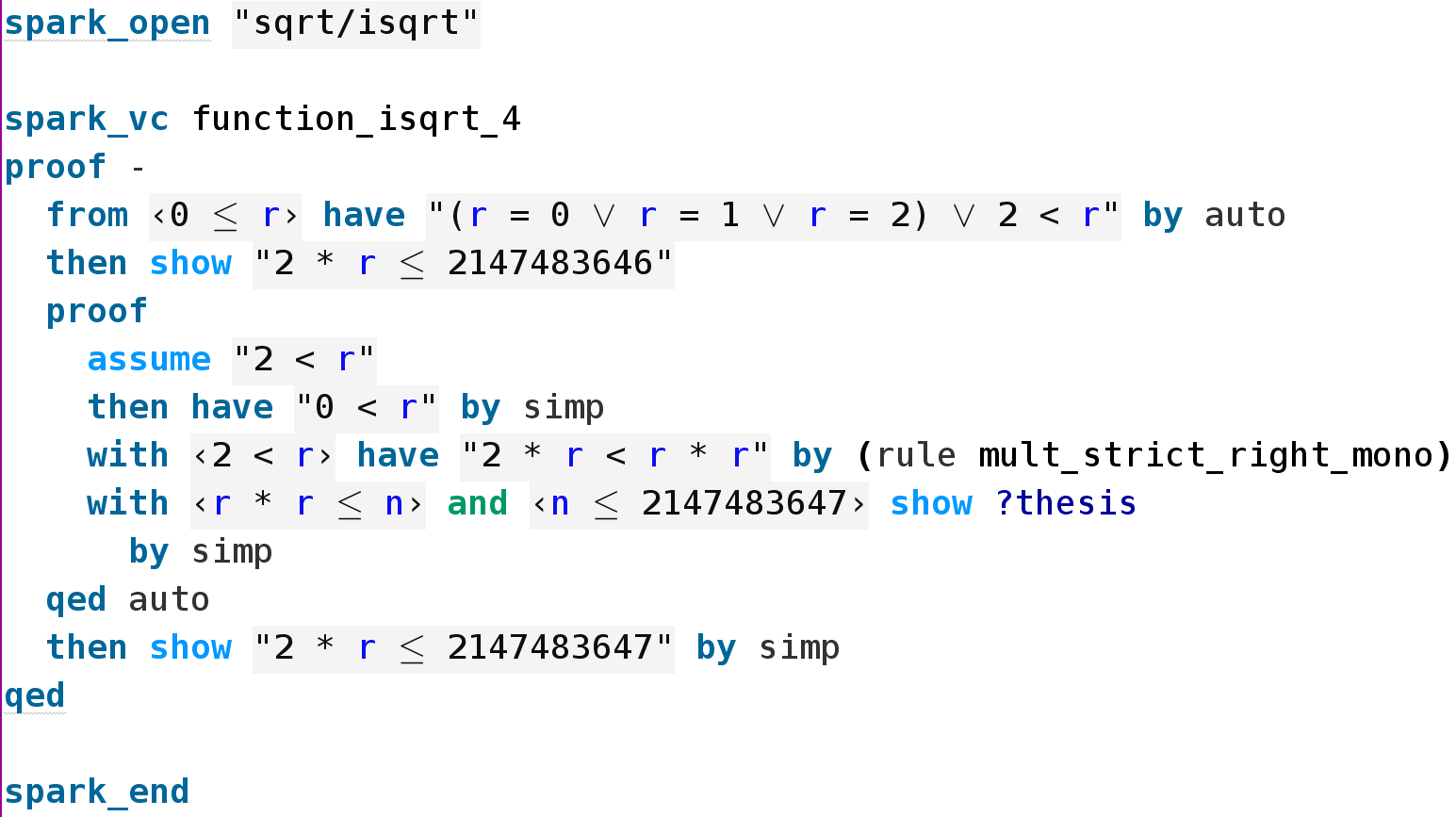}
  \caption{Isabelle/HOL-SPARK importing and proving a verification condition}
  \label{fig:spark}
  \end{figure}

  E.g. consider \isatt{{\char`\$}ISABELLE{\char`\_}HOME/\discretionary{}{}{}src/\discretionary{}{}{}HOL/\discretionary{}{}{}SPARK/\discretionary{}{}{}Examples/\discretionary{}{}{}Sqrt/\discretionary{}{}{}Sqrt.thy}
  (\figref{fig:spark}): it uses the load command \textbf{spark\_open} to read
  verification conditions from the output files of other SPARK tools by Altran
  Praxis Ltd (the results are already included in the Isabelle distribution).
  This augments the theory data slots defined by HOL-SPARK, such that other
  commands can refer to it in the subsequent document, notably \textbf{spark\_vc} to
  retrieve a named verification condition and commence a proof. The final
  \textbf{spark\_end} command ensures that all pending VCs have been solved.

  It is easy to inspect the command implementations, by using
  control-hover-click on the keywords in the text, which leads to
  \isatt{{\char`\$}ISABELLE{\char`\_}HOME/\discretionary{}{}{}src/\discretionary{}{}{}HOL/\discretionary{}{}{}SPARK/\discretionary{}{}{}Tools/\discretionary{}{}{}spark{\char`\_}commands.ML} and its auxiliary
  ML definitions to implement command parsers and theory data management. Thus
  we see how tool development and tool usage happens within the same universal
  IDE.%
\end{isamarkuptext}\isamarkuptrue%
\isadelimdocument
\endisadelimdocument
\isatagdocument
\isamarkupsection{Document Text Structure \label{sec:document-text}%
}
\isamarkuptrue%
\endisatagdocument
{\isafolddocument}%
\isadelimdocument
\endisadelimdocument
\begin{isamarkuptext}%
PIDE documents may contain embedded formal languages, together with
  \emph{informal text}, which vaguely resembles {\LaTeX} source. The structure of
  informal text is explicitly specified, as document \emph{markup commands} (e.g.
  ``\isa{\isacommand{section}\ {\isacartoucheopen}text{\isacartoucheclose}}'' or ``\isa{\isacommand{text}\ {\isacartoucheopen}text{\isacartoucheclose}}''), and a version of \emph{markdown
  notation} for item lists within the \isa{text} body. Moreover, various
  sub-languages of Isabelle admit \emph{marginal comments} alongside the formal
  content (e.g. ``\isa{{\isasymcomment}}~\isa{{\isasymopen}text{\isasymclose}}''). Any such informal text may again refer to
  formal languages via \emph{document antiquotations}, (e.g. ``\isa{{\isacharat}{\isacharbraceleft}term\ {\isasymopen}t{\isasymclose}{\isacharbraceright}}'').

  Historically, some programming languages (e.g. Java) have introduced
  enhanced source comments by detecting special ``doc comments'' via add-on
  tools (e.g. \isatt{javadoc}), but Isabelle document text goes beyond this: it has
  been granted first-class status within the syntax. Note that Isabelle
  supports \emph{source comments} as well (e.g. \isatt{(*\ source\ *)}), but these are
  omitted in document processing (like \isatt{{\char`\%}} in {\LaTeX}).\footnote{Users sometimes
  misunderstand Isabelle theory sources as ``program code'' and consequently
  use old-fashioned \isatt{(*\ source\ *)} comments to explain what they write, but
  readers of the final document won't see that: it is stripped from the input
  and thus absent in the output.}

  Below follows a systematic overview of Isabelle language elements for
  document text structure, with fine points of notation and rendering in
  Isabelle/jEdit.%
\end{isamarkuptext}\isamarkuptrue%
\isadelimdocument
\endisadelimdocument
\isatagdocument
\isamarkupparagraph{Markup commands%
}
\isamarkuptrue%
\endisatagdocument
{\isafolddocument}%
\isadelimdocument
\endisadelimdocument
\begin{isamarkuptext}%
are special theory commands with a single \isa{text} argument (double-quoted
string or cartouche). Formally, this is the identity function on theory
content, but the argument text is checked within the context: it may contain
antiquotations. Informally, markup commands have a meaning to Isabelle
document preparation, as section headings or paragraphs of plain text.

The following markup commands are predefined in the Isabelle/Pure bootstrap,
and cannot be extended in user-space (which is atypical for Isabelle):

\begin{itemize}%
\item Section headings (6 levels as in HTML): \isa{\isacommand{chapter}}, \isa{\isacommand{section}},
\isa{\isacommand{subsection}}, \dots, \isa{\isacommand{subparagraph}}. Section labels are not treated
formally, but can be used via raw {\LaTeX}, e.g.
``\isa{\isacommand{section}\ {\isacartoucheopen}Foo\ bar\ bla\ {\isacharbackslash}label{\isacharbraceleft}sec{\isacharcolon}foo{\isacharbraceright}{\isacartoucheclose}}''.

Isabelle/jEdit turns section headings into a tree-view in the \emph{SideKick}
panel: this provides an informal document outline, over the structure of
formal elements (e.g. definitions, theorem statements, while omitting
proofs).

\item Text blocks via \isa{\isacommand{text}} (paragraph with standard style), \isa{\isacommand{txt}}
(paragraph with slightly smaller style), \isa{\isacommand{text{\isacharunderscore}raw}} (no change of style,
e.g. for raw {\LaTeX}).

Plain words within the text are marked up for spell-checking in the IDE:
words that are missing from the (English) dictionary are rendered with
blue underline, to say gently that something might be wrong. That markup
is also used by the completion mechanism, to propose alternative words
from the dictionary. For example, in \figref{fig:rail} the word
``criterium'' is actually wrong, but the technical term ``Isar'' is just
unknown in the dictionary: it can be amended by suitable actions in the
right-click menu of jEdit.%
\end{itemize}%
\end{isamarkuptext}\isamarkuptrue%
\isadelimdocument
\endisadelimdocument
\isatagdocument
\isamarkupparagraph{Markdown lists%
}
\isamarkuptrue%
\endisatagdocument
{\isafolddocument}%
\isadelimdocument
\endisadelimdocument
\begin{isamarkuptext}%
may occur within the body of text blocks, notably for commands \isa{\isacommand{text}} and
\isa{\isacommand{txt}}. This quasi-visual format resembles official
\emph{Markdown}\footnote{\url{http://commonmark.org}}, but the full complexity of that
notation is avoided. Instead of ASCII art, we use the following Isabelle
symbols as markers for list items: \isatt{{\char`\\}{\char`\<}{\char`\^}item{\char`\>}} for itemize, \isatt{{\char`\\}{\char`\<}{\char`\^}enum{\char`\>}} for enumerate,
\isatt{{\char`\\}{\char`\<}{\char`\^}descr{\char`\>}} for description. Adjacent list items with same indentation and same
marker are grouped into a single list. Singleton blank lines separate
paragraphs. Multiple blank lines escape from the current list hierarchy.

\begin{figure}[h!]
\centering
\includegraphics[width=0.4\textwidth]{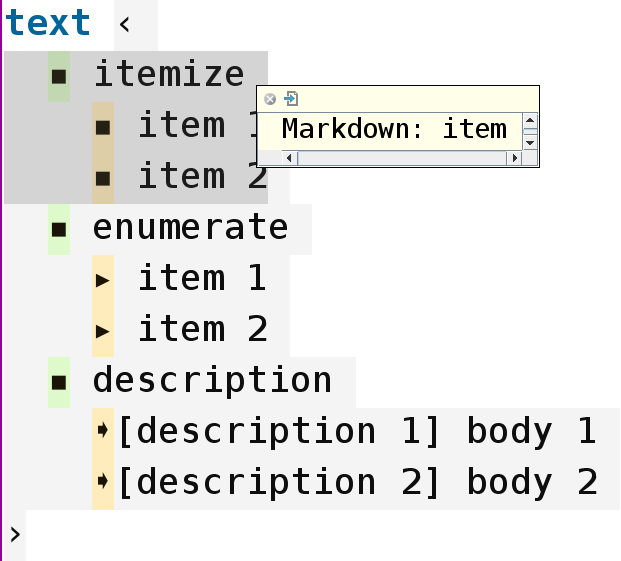}
\caption{Markdown notation for lists in Isabelle/jEdit}
\label{fig:markdown}
\end{figure}

Isabelle/jEdit renders item markers nicely, using a special font. The depth
of nested lists is emphasized by the standard color scheme for \emph{text folds}
in jEdit. Lists, items, and constituent text paragraphs are marked-up in the
PIDE document model, such that control-hovering with the mouse reveals that
structure to the user; see also \figref{fig:markdown}.

Isabelle users have occasionally reported that they like this enhanced
source representation better than classic WYSIWYG text processors, because
the structure of nested list items is clearly seen and easily edited.%
\end{isamarkuptext}\isamarkuptrue%
\isadelimdocument
\endisadelimdocument
\isatagdocument
\isamarkupparagraph{Formal comments%
}
\isamarkuptrue%
\endisatagdocument
{\isafolddocument}%
\isadelimdocument
\endisadelimdocument
\begin{isamarkuptext}%
are an integral part of the document, but are logically void and removed
from the resulting theory or term content. Document output supports various
text styles, according to the subsequent kinds of comments.

\begin{itemize}%
\item Marginal comment of the form ``\isa{{\isasymcomment}}~\isa{{\isasymopen}text{\isasymclose}}'' (literally
``\isatt{{\char`\\}{\char`\<}comment{\char`\>}}~\isa{{\isasymopen}text{\isasymclose}}''). The given argument is typeset as regular text, with
formal antiquotations in the body. This allows to alternate formal
sub-languages and informal text arbitrarily. Input works e.g. by
completion of ``\isatt{{\char`\\}co}''.

\item Canceled text of the form ``\isa{\isactrlcancel }\isa{{\isasymopen}text{\isasymclose}}'' (literally ``\isatt{{\char`\\}{\char`\<}{\char`\^}cancel{\char`\>}}\isa{{\isasymopen}text{\isasymclose}}'').
The argument is typeset as formal Isabelle source and overlaid with a
``strike-through'' pattern, e.g. \isa{%
\isamarkupcancel{bad}}. Input works e.g. by completion
of ``\isatt{{\char`\\}ca}''.

\item {\LaTeX} source of the form ``\isa{\isactrllatex }\isa{{\isasymopen}text{\isasymclose}}'' (literally
``\isatt{{\char`\\}{\char`\<}{\char`\^}latex{\char`\>}}\isa{{\isasymopen}text{\isasymclose}}''). This allows to emit arbitrary (unchecked)
{\LaTeX} source. Input works e.g. by completion of ``\isatt{{\char`\\}la}''.%
\end{itemize}

These formal comments work uniformly in the theory and term language, but
also in Isabelle/ML and some other embedded languages. User-defined
languages can easily include formal comments via standard scanner functions
provided in module \isatt{Comment} of Isabelle/ML and Isabelle/Scala. A clash
with existing language syntax is unlikely, due to the peculiar notation
with non-ASCII Isabelle symbols.%
\end{isamarkuptext}\isamarkuptrue%
\isadelimdocument
\endisadelimdocument
\isatagdocument
\isamarkupparagraph{Document antiquotations%
}
\isamarkuptrue%
\endisatagdocument
{\isafolddocument}%
\isadelimdocument
\endisadelimdocument
\begin{isamarkuptext}%
provide a uniform scheme to embed domain-specific formal languages within
Isabelle document text (even some other sub-languages). The terminology is
explained as follows: the outer formal syntax allows to \emph{quote} informal
text, which in turn allows to \emph{antiquote} formal content. E.g. the term
antiquotation ``\isa{{\isacharat}{\isacharbraceleft}term\ {\isasymopen}Groups{\isachardot}plus\ x\ y{\isasymclose}{\isacharbraceright}}'' pretty-prints its argument
into the generated PDF-{\LaTeX} document using the concrete syntax from the
theory context for the constant \isa{Groups{\isachardot}plus}, it becomes
``\isa{x\ {\isacharplus}\ y}''. It is also possible to force printing of the
original source text of the (formally checked) term, so \isa{{\isacharat}{\isacharbraceleft}term\ {\isacharbrackleft}source{\isacharbrackright}\ {\isasymopen}Groups{\isachardot}plus\ x\ y{\isasymclose}{\isacharbraceright}} becomes ``\isa{{\isasymopen}Groups{\isachardot}plus\ x\ y{\isasymclose}}''.

\medskip Antiquotation syntax comes in the following variants:

\begin{description}%
\item [Long form:] \isatt{@{\char`\{}}\isa{name\ {\isacharbrackleft}options{\isacharbrackright}\ arguments\ {\isasymdots}}\isatt{{\char`\}}} with free-form
arguments according to the parser provided for this antiquotation.

\item [Short forms:] ~

\begin{enumerate}%
\item \isatt{{\char`\\}}\isatt{{\char`\<}{\char`\^}}\isa{name}\isatt{{\char`\>}}\isa{{\isasymopen}argument{\isasymclose}} for \isatt{@{\char`\{}}\isa{name\ {\isasymopen}argument{\isasymclose}}\isatt{{\char`\}}}, i.e.
there are no options and exactly one argument that is a cartouche.

\item \isa{{\isasymopen}argument{\isasymclose}} for \isatt{@{\char`\{}}\isa{cartouche\ {\isasymopen}argument{\isasymclose}}\isatt{{\char`\}}}, i.e. as above, but
the name is the special constant ``\isa{cartouche}''. It means that one
sub-language in the current context can get the privilege to be embedded
without explicit marker.

\item \isatt{{\char`\\}}\isatt{{\char`\<}{\char`\^}}\isa{name}\isatt{{\char`\>}} for \isatt{@{\char`\{}}\isa{name}\isatt{{\char`\}}} without options or arguments.%
\end{enumerate}%
\end{description}

An entity of the form \isatt{{\char`\\}}\isatt{{\char`\<}{\char`\^}}\isa{name}\isatt{{\char`\>}} is called \emph{control symbol} in
Isabelle: here it acts like an operator on the subsequent text cartouche.
Isabelle/jEdit provides two ways to render control symbols nicely:

\begin{enumerate}%
\item As Unicode glyphs, according to a global symbol table of the Isabelle
distribution. E.g. \isatt{{\char`\\}{\char`\<}{\char`\^}file{\char`\>}}, \isatt{{\char`\\}{\char`\<}{\char`\^}dir{\char`\>}}, \isatt{{\char`\\}{\char`\<}{\char`\^}url{\char`\>}} are rendered as icons that are
similar to common desktop file-managers.

\item As bold-italic keyword (default). For example, \isatt{{\char`\\}{\char`\<}{\char`\^}term{\char`\>}} becomes
\isa{\isactrlterm }, and \isatt{{\char`\\}{\char`\<}{\char`\^}prop{\char`\>}} becomes \isa{\isactrlprop }. This form serves as
fall-back for names that lack a clear visual representation, and for
arbitrary control symbols that users might invent for their own
applications.%
\end{enumerate}

\medskip Rendering of antiquotation arguments follows the standard mechanisms of
the Prover IDE, e.g. a term argument is decorated with colors for free and
bound variables, identifier scopes, inferred types, hyper links etc. For
unusual arguments, such as system resources, the semantic operation of the
antiquotation emits markup to instruct the front-end to do the right thing,
e.g. open a file in the text editor, or a directory in the file-system
browser, or a URL in an external web browser (see
\figref{fig:antiquote-resources}).

\begin{figure}[h!]
\centering
\includegraphics[width=0.95\textwidth]{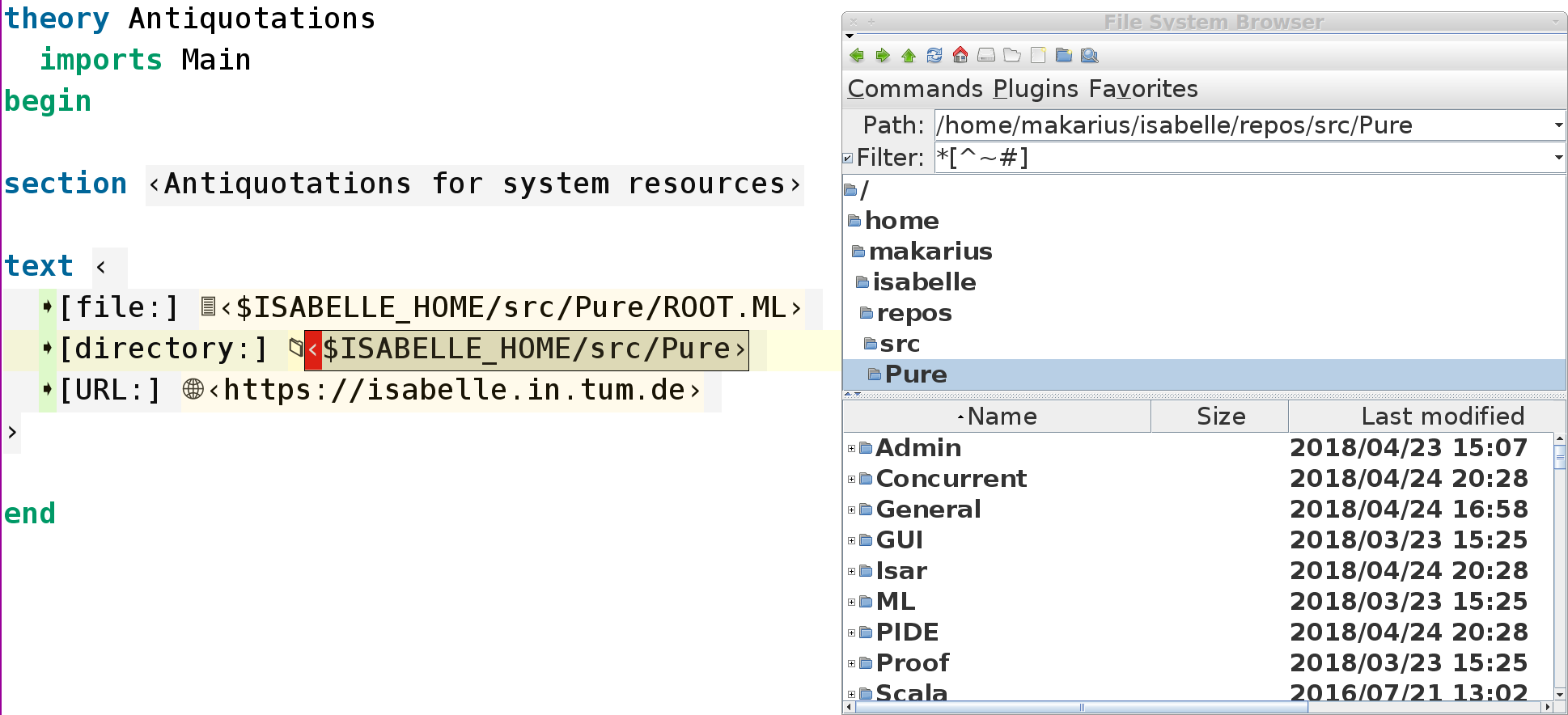}
\caption{Document antiquotations for system resources (action on directory)}
\label{fig:antiquote-resources}
\end{figure}

Isabelle tool developers may figure out how this is implemented, by putting
an example antiquotation into the document, and performing
control-hover-click onto its name to get to the definition in Isabelle/ML,
and then use the ML IDE to explore that further (that also requires to open
\isatt{{\char`\$}ISABELLE{\char`\_}HOME/\discretionary{}{}{}src/\discretionary{}{}{}Pure/\discretionary{}{}{}ROOT.ML}). This principle of self-application and
self-sufficiency of Isabelle/jEdit makes it a truly integrated development
environment.

\bigskip \Figref{fig:nesting} shows a somewhat artificial example for the nesting
of sub-languages within a proof document.\footnote{See also
  \url{https://bitbucket.org/makarius/fide2018/raw/tip/Nesting.thy} and recall
  that Isabelle/jEdit can open URLs of text files directly.} It builds a local
context with fixed variables and assumptions, followed by a \isa{\isacommand{text}} block
with embedded \isa{{\isacharat}{\isacharbraceleft}lemma{\isacharbraceright}} antiquotations: the stated results are proven in
that context, but not recorded as facts. The final PDF merely prints the
lemma statements as plain propositions.

\begin{figure}[h!]
\centering
\includegraphics[width=0.95\textwidth]{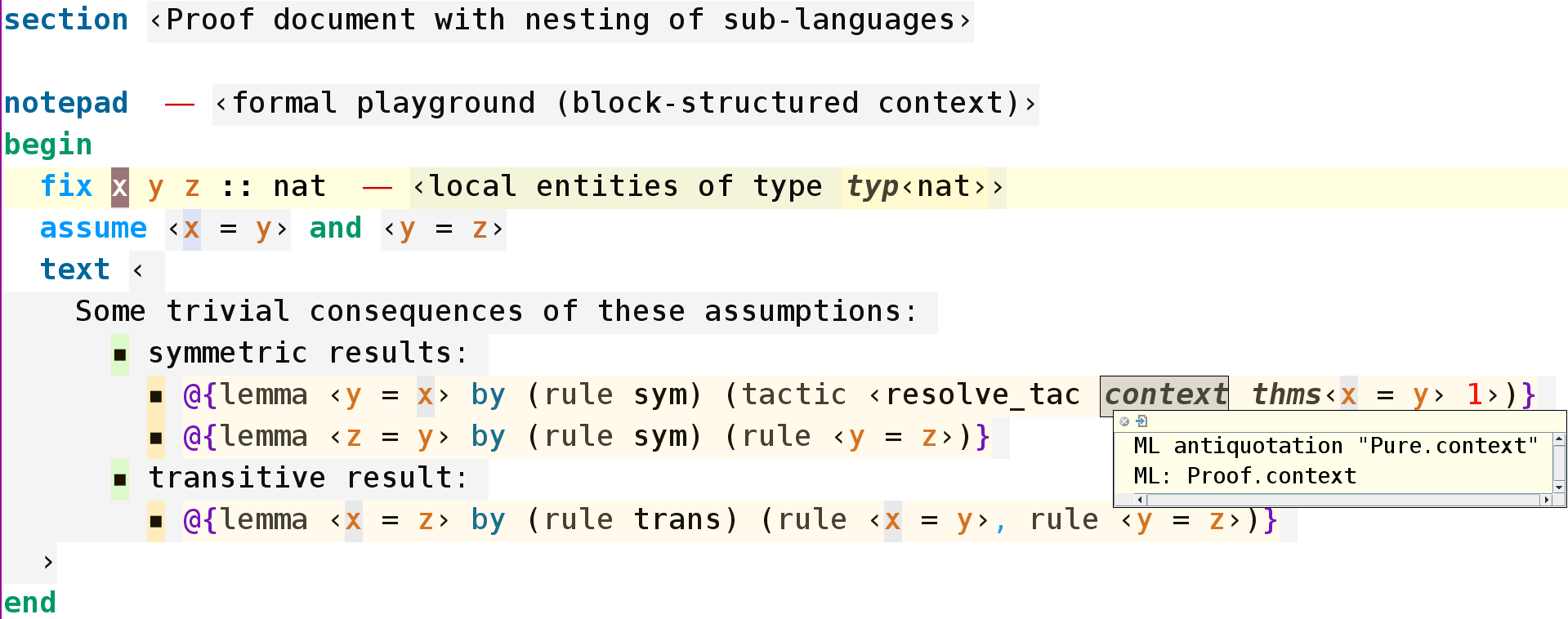}
\caption{Proof document with nesting of sub-languages}
\label{fig:nesting}
\end{figure}

Here the cursor has been placed at the binding position of the variable \isa{x}:
its subsequent uses are highlighted, thanks to markup provided by the
sub-language parsers. It is also possible turn that scope group into a
multi-selection of the editor, and thus rename the variable systematically
(but without any sanity checks). The information about occurrences of \isa{x} is
produced by the individual parsers for nested antiquotations, which in turn
refer to standard parsers of a term in a certain context: that emits PIDE
markup for constants and variables encountered in the nested text.

The proof of \isa{{\isacharat}{\isacharbraceleft}lemma\ {\isasymopen}y\ {\isacharequal}\ x{\isasymclose}{\isacharbraceright}} uses the proof method \isa{tactic},
which takes an ML expression (of a suitable type) as argument. The
highlighted box over \isa{\isactrlcontext } is the result of a control-hover GUI
action; it explains that spot formally in the tooltip.

\medskip
Thus users can explore the formal structure of example documents, and learn
something about syntax and semantics of the many domain-specific formal
languages of Isabelle, even without studying the documentation.%
\end{isamarkuptext}\isamarkuptrue%
\isadelimdocument
\endisadelimdocument
\isatagdocument
\isamarkupsection{Example: BibTeX Database and Citation Management \label{sec:bibtex}%
}
\isamarkuptrue%
\endisatagdocument
{\isafolddocument}%
\isadelimdocument
\endisadelimdocument
\begin{isamarkuptext}%
Management of bibliographies and citations is definitely outside of formal
logic and functional programming, but it is practically relevant for
Isabelle document authoring. Isabelle/jEdit provides IDE support as
illustrated in \figref{fig:bibtex}, with the following features:

\begin{itemize}%
\item A jEdit syntax mode for \isatt{.bib} files: it recognizes the BibTeX syntax
according to the original parser written in Pascal, re-implemented in
Isabelle/Scala.\footnote{See also
    \url{http://ctan.org/tex-archive/biblio/bibtex/base/bibtex.web} module
    \isatt{@{\char`\<}Scan\ for\ and\ process\ a\ {\char`\\}.{\char`\{}.bib{\char`\}}\ command\ or\ database\ entry@{\char`\>}}.} It
recognizes the token language, and the names and block structure of
entries.

\item Support for text folds according to the block structure of BibTeX
entries.

\item A tree-view in the \isatt{SideKick} panel: each entry is associated with a
name and content. The jEdit \isatt{SideKick} plugin allows to filter the
tree-view by giving a substring for the name.

\item A context-menu for BibTeX entry types (e.g. \emph{Article},
\emph{InProceedings}).

\item Syntax highlighting for BibTeX entry fields, depending on the entry
type, to distinguish required vs. optional fields.

\item HTML preview similar to {\LaTeX} output, using the \isatt{bib2xhtml}
tool.\footnote{\url{https://github.com/dspinellis/bib2xhtml}} Users need to invoke
the jEdit action \isatt{isabelle.preview} on an open \isatt{.bib} buffer.

\item Semantic checking by the original \isatt{bibtex} tool (using a default style
file). This attaches warnings and errors to the source, while the user
is editing.

\item Support for document antiquotations \isa{{\isacharat}{\isacharbraceleft}cite\ NAME{\isacharbraceright}}, with name completion
according to all \isatt{.bib} files that happen to be open in the editor.

\item Strict checking of cited BibTeX entries wrt. the \isatt{.bib} files in
\textbf{document\_files} of the enclosing session; this only works for
batch-builds, e.g. when producing the final PDF.%
\end{itemize}

\begin{figure}[h!]
\centering
\includegraphics[width=0.95\textwidth]{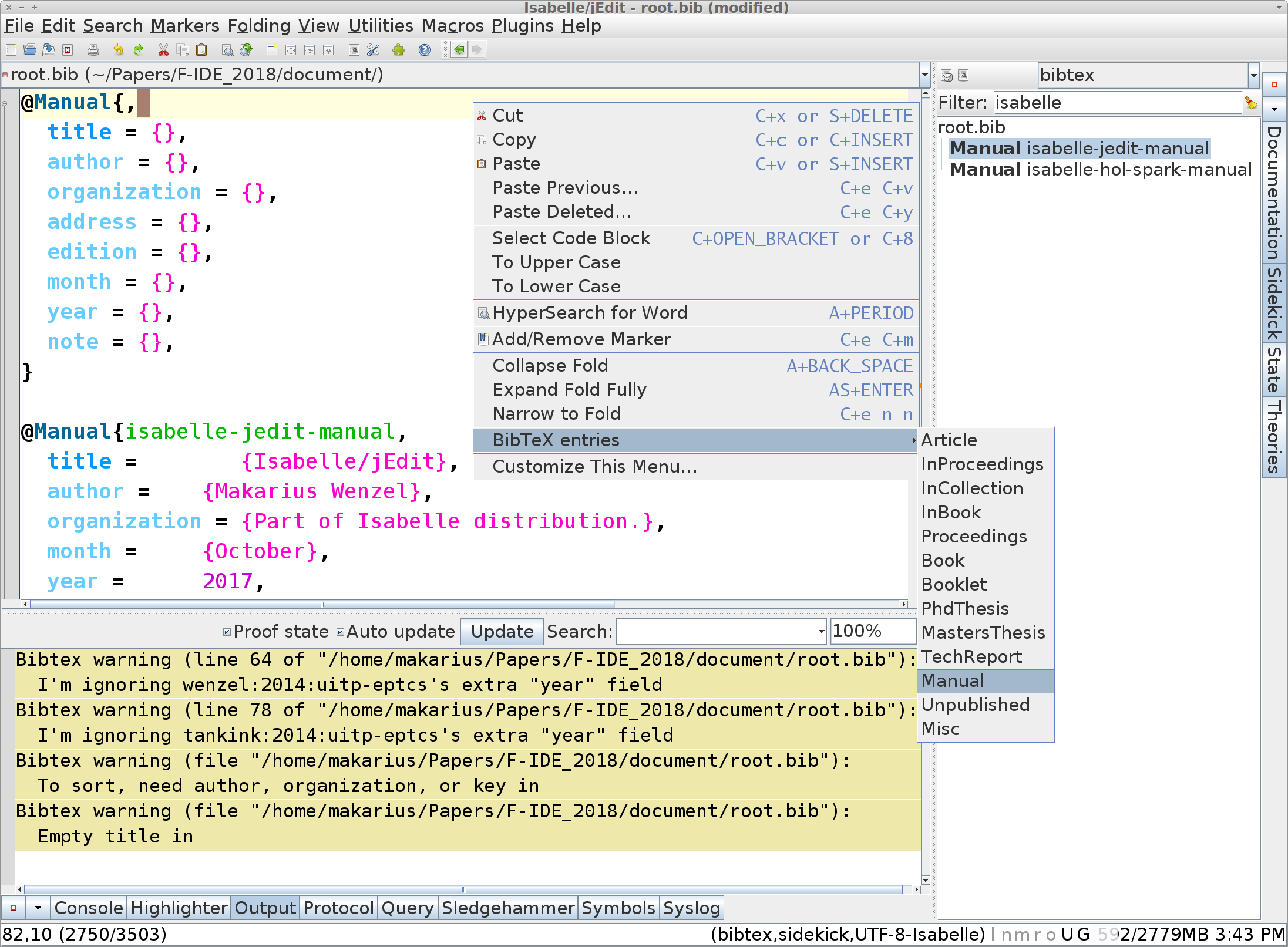}
\caption{Semantic IDE support for BibTeX databases}
\label{fig:bibtex}
\end{figure}

The implementation of all this is quite concise. It mainly happens in
Isabelle/Scala\footnote{See \isatt{{\char`\$}ISABELLE{\char`\_}HOME/\discretionary{}{}{}src/\discretionary{}{}{}Pure/\discretionary{}{}{}Thy/\discretionary{}{}{}bibtex.scala} and
  \isatt{{\char`\$}ISABELLE{\char`\_}HOME/\discretionary{}{}{}src/\discretionary{}{}{}Tools/\discretionary{}{}{}jEdit/\discretionary{}{}{}src/\discretionary{}{}{}jedit{\char`\_}bibtex.scala}.}, but semantic
processing of \isatt{.bib} files and the \isa{{\isacharat}{\isacharbraceleft}cite{\isacharbraceright}} antiquotation need to happen
in the stateless mathematical world of Isabelle/ML: special PIDE protocol
messages are used to invoke a Scala function \isatt{String\ ={\char`\>}\ String} and return
its result in ML. Some further tricks are required to process \isatt{.bib} files
faithfully and to produce messages with proper positions:

\begin{itemize}%
\item Opening the file \isatt{foo.bib} in Isabelle/jEdit implicitly puts it into a
theory context (derived from \isatt{Pure}) with the load command \isa{\isacommand{bibtex{\isacharunderscore}file}\ {\isacartoucheopen}foo{\isachardot}bib{\isacartoucheclose}} in the body. This means that \isatt{.bib} files automatically become
auxiliary files as explained in \secref{sec:document-structure}.

\item The command \isa{\isacommand{bibtex{\isacharunderscore}file}} is implemented in Isabelle/ML as a function
\isatt{string\ {\char`\-}{\char`\>}\ unit} that emits PIDE messages according to its true
meaning (here only warnings and errors). This works by asking
Isabelle/Scala to invoke \isatt{Bibtex.check{\char`\_}database()} to do the main job.

\item The Scala function \isatt{Bibtex.check{\char`\_}database()} tokenizes the given source
text with the Isabelle BibTeX parser, which gives precise position
information. Each token is put on a line of its own and the result given
to the \isatt{bibtex} executable as temporary file, with options to use the
\isatt{plain.bst} style file. Note that BibTeX styles are programs in a special
language, to analyse and output database entries.

\item The resulting BibTeX log file is scanned for warnings and errors, while
observing its peculiar format (following the original implementation in
Pascal). Line numbers are used as index positions for the tokens produced
by the Isabelle BibTeX parser: the final results are precise positions for
PIDE messages, which users will see as squiggly underlines in the source.%
\end{itemize}

Thus we can pretend that Isabelle understands the meaning of BibTeX files,
or that BibTeX understands the PIDE protocol. The Scala and ML sources for
that are less complex than the above explanations might suggest. It
demonstrates that unusual IDE applications work out with quite reasonable
effort.%
\end{isamarkuptext}\isamarkuptrue%
\isadelimdocument
\endisadelimdocument
\isatagdocument
\isamarkupsection{Conclusion%
}
\isamarkuptrue%
\isamarkupsubsection{PIDE History and Related Work%
}
\isamarkuptrue%
\endisatagdocument
{\isafolddocument}%
\isadelimdocument
\endisadelimdocument
\begin{isamarkuptext}%
Isabelle once shared the well-known Proof General Emacs interface \cite{Aspinall-et-al:2007} with other proof assistants, most notably Coq. In
  2008, I started to think beyond the Proof General model of stepping forwards
  and backwards through a ``script'' of prover commands. This eventually lead
  to the PIDE document model and the Isabelle/jEdit front-end, all implemented
  in Isabelle/Scala. The new prover front-end was sufficiently
  well-established in 2014 to remove the last remains of the command loop, so
  that Proof General is now exclusively used for Coq and has de-facto lost its
  generality.\footnote{\url{https://proofgeneral.github.io}}

  PIDE is document-oriented and timeless / stateless: it operates via
  functional updates on explicitly identified versions. Proof General and its
  many derivatives are script-oriented and stateful: one command after another
  on a global version. This means that various Proof General upgrades and
  clones (e.g. CoqIde or Coqoon\footnote{\url{https://coqoon.github.io}}) are actually
  \emph{unrelated} to PIDE. An exception is the Coq PIDE experiment from 2013/2014
  \cite{Tankink:2014:UITP-EPTCS}, but it did not reach end-users so far.

  Mainstream IDEs (e.g. Eclipse, Netbeans, IntelliJ IDEA) are more closely
  related to PIDE, but there is a conceptual difference: classic IDEs operate
  by static analysis of the program source and provide a separate debugger for
  dynamic exploration at runtime. In contrast, the PIDE model explores the
  syntactic structure and semantic content of the document uniformly: this
  works, because semantic operations are usually pure, without side-effects.
  Isabelle/jEdit also provides a classic debugger for Isabelle/ML, but that is
  strictly speaking not part of the PIDE document model: it interacts with the
  running ML program on a side-channel and a separate GUI panel.

  Visual Studio Code (VSCode) is a remarkable open-source project by
  Microsoft\footnote{\url{https://code.visualstudio.com}}. Instead of a full-blown IDE,
  it is positioned as sophisticated a text editor, with support for ``language
  smartness'' provided by VSCode extensions (e.g. for JavaScript, TypeScript,
  Java). There is also a published \emph{Language Server Protocol} to connect an
  external process that turns edits of sources into reports about its
  structure and meaning. This idea is very close to Isabelle/PIDE, and I have
  already connected the Isabelle/Scala process as external ``language
  smartness provider'' for Isabelle theory and ML files.

  The resulting Isabelle/VSCode 1.1 for Isabelle2018 basically
  works,\footnote{\url{https://marketplace.visualstudio.com/items?itemName=makarius.Isabelle2018}}
  but has a long way ahead, in order to catch up with Isabelle/jEdit 10.0 in
  that release. This is not just a matter to improve the Isabelle side: VSCode
  is quite different from jEdit in many respects. For example, VSCode imposes
  more explicit structure and policies on the internal text model, while jEdit
  is just a multi-layered paint program (for Java2D) with a plain text buffer
  behind it. Thus the Isabelle/jEdit ``game engine'' for real-time text
  rendering with rich markup was relatively easy to implement, by removing the
  default text painting layers and adding specific ones for PIDE. For VSCode,
  more work will be required to take it apart and access its HTML/CSS
  document-model directly (even worse, Microsoft explicitly restricts the
  access of external extensions to these important internals).

  Nonetheless, the VSCode platform looks like an interesting alternative for
  future versions of the Isabelle Prover IDE. Some other proof assistants have
  already started to support it:
  VScoq\footnote{\url{https://github.com/siegebell/vscoq}} as a Proof General clone, and
  VSCode-Lean\footnote{\url{https://github.com/leanprover/vscode-lean}} based on a
  custom-made theory compilation server for Lean Prover \cite{Moura-et-al:2015}.%
\end{isamarkuptext}\isamarkuptrue%
\isadelimdocument
\endisadelimdocument
\isatagdocument
\isamarkupsubsection{Future Work%
}
\isamarkuptrue%
\endisatagdocument
{\isafolddocument}%
\isadelimdocument
\endisadelimdocument
\begin{isamarkuptext}%
Even after 10 years, the Isabelle/PIDE framework and the Isabelle/jEdit
front-end are still not finished, but an open-ended enterprise. Here are
possible improvements only for the topics covered in this paper:

\begin{itemize}%
\item More explicit support for session definitions in the IDE. So far,
Isabelle/jEdit uses the union of all sessions from the active project
directories, and is thus able to map theory files to logical theory names.
It would be nice to take specific \textbf{options} and \textbf{document\_files} from
session ROOT files into account. The latter would also allow to relate
BibTeX files to theories implicitly, and thus provide a context for
\isa{{\isacharat}{\isacharbraceleft}cite{\isacharbraceright}} antiquotations, without asking the user to open relevant \isatt{.bib}
files.

\item Building PDF-{\LaTeX} documents on the spot within the IDE. So far this
is still a batch-job, either via \isatt{isabelle\ build} on the shell, or via
\isatt{Build.build()} in the Scala console of Isabelle/jEdit. Full integration
into the IDE would mean that the user can select relevant theories for PDF
output in a GUI panel, push a button, and see the generated (or updated)
PDF quickly, lets say within 1--3s. In order to achieve this, the Isabelle
document preparation system needs to be removed from the traditional
Isabelle/ML session build process, and integrated into Isabelle/Scala.
This reorganization would also grant access to PIDE markup information,
such that the generated PDF could become more detailed (e.g. different
text style for different identifiers: free variables vs. bound variables
vs. constants).

\item High-quality HTML document output, as a continuation of existing
PDF-{\LaTeX} output and the IDE rendering of the theory source. HTML + CSS
have become sufficiently powerful for high-quality type-setting of books
and journals. Imitating PDF-{\LaTeX} documents in HTML would also spare
time-consuming invocations of \isatt{pdflatex}, and thus make document output a
matter of milliseconds instead of seconds. In recent years, we have seen
text editors with Markdown source vs. preview in a split window: something
like that could be achieved for Isabelle HTML documents as well, but with
rich semantic PIDE markup. Instantaneous document preview might also help
to overcome the occasional misunderstanding of users, to think of theory
sources as ``program code'' instead of a document.%
\end{itemize}

Apart from further refinement of the PIDE technology, we need to work more
on the \emph{sociology} of the project: Isabelle/jEdit is hardly known outside
of the Isabelle community, and inside it users merely take it for granted.
Hopefully this paper helps to apply PIDE in other applications, by Isabelle
users or people building different formal tools.%
\end{isamarkuptext}\isamarkuptrue%
\isadelimtheory
\endisadelimtheory
\isatagtheory
\isacommand{end}\isamarkupfalse%
\endisatagtheory
{\isafoldtheory}%
\isadelimtheory
\endisadelimtheory
\end{isabellebody}%